\documentclass[useAMS,usenatbib]{mn2e}
\usepackage{psfig}

\newcommand{\be}{\begin{equation}}
\newcommand{\en}{\end{equation}}
 
\newcommand{\lapp}{\mbox{\raisebox{-0.3em}{$\stackrel{\textstyle <}{\sim}$}}}
\newcommand{\gapp}{\mbox{\raisebox{-0.3em}{$\stackrel{\textstyle >}{\sim}$}}}

\title[Unification scheme and neutral gas]{Unification scheme and the distribution of neutral gas
in compact radio sources}
\author[Neeraj Gupta and D.J. Saikia]{Neeraj Gupta$^1$\thanks{E-mail:neeraj@ncra.tifr.res.in (NG); 
djs@ncra.tifr.res.in (DJS)} and D.J. Saikia$^{1,2}$ \\
$^1$ National Centre for Radio Astrophysics, TIFR, Pune 411 007, India \\
$^2$ Jodrell Bank Observatory, University of Manchester, Macclesfield, Cheshire SK11 9DL} 
\begin{document}

\date{Accepted. Received; in original form }

\pagerange{\pageref{firstpage}--\pageref{lastpage}} \pubyear{2005}

\maketitle

\label{firstpage}

\begin{abstract}
We examine the consistency of H{\sc i} properties with the unification scheme for
radio galaxies and quasars, and any correlation
with the symmetry parameters for a sample of CSS and GPS sources. In our sample, 15 out of 23
galaxies exhibit 21-cm absorption as against 1 out of 9 quasars, which is broadly 
consistent with the unification scheme. 
Also there is a tendency for the detection rate as well as the column density for galaxies 
to increase with core prominence, f$_c$, a statistical indicator of the orientation of the jet 
axis to the line of sight.
This can be understood in a scenario where radio sources are larger than the scale of the circumnuclear 
H{\sc i} disk so that the lines of sight to the lobes at very large inclinations do not intersect the disk.  
The sources 
in our sample also exhibit the known anti-correlation between H{\sc i} column density and source size.  
This suggests that small linear size, along with intermediate values of core prominence, is a good recipe 
for detecting 21-cm absorption in CSS and GPS objects.  
Some of the absorption could also be arising from clouds which may have interacted with the radio jet.  
The H{\sc i} column density and velocity shift of the primary absorption component, however, show no 
dependence on the degree of misalignment and the separation ratio of the radio sources.    
\end{abstract}

%
\begin{table*}
\caption{Sample of CSS and GPS sources.}
\begin{center}
\begin{tabular}{|l|l|c|l|l|l|r|l|l|l|l|l|r|r|l|}
\hline
Source & Alt. & Opt. & Redshift & P$_{\rm 5GHz}$        &  LLS   &f$_{c}$& Str.  & R$_\theta$ & $\Delta$ &Ref. & Radio & N(H{\sc i}) & V$_{shift}$ & Ref. \\
name   & name & ID   &          & $\times$10$^{25}$ &        &       & class &     &   &   & class & $\times$10$^{20}$ \\
       &      &      &          &  (W/Hz)           & (kpc)  &       &       &   &  ($^\circ$) &    &       &  (cm$^{-2}$) &  (km/s) & \\ 
 (1)   &  (2) &  (3) &   (4)    &  (5)      & (6)    & (7)   &  (8)  &(9)&  (10) & (11)&  (12)  &  (13) &  (14) & (15)  \\
\hline                                                                        
J0025$-$2602& OB-238 & G & 0.322  &  112     & 3.02      &$<$0.002  & D  &    &     &1    & CSS   &   2.42 &$-$30  & V \\ 
J0111+3906  & OC314  & G & 0.669  &  99.6    & 0.04      &$<$0.004  & T  &1.28&9.7  &2,3  & GPS   &   84.2 &   0   & C\\
            &        &   &        &          &           &          &    &    &     &4,5  &       &        &       &   \\ 
J0119+3210  & 4C31.04& G & 0.060  &  1.30    & 0.09      &   0.012  & T  &1.14&7.0  &6    & CSS   &   12.2 &$-$10  & Gu\\
J0137+3309  & 3C48   & Q & 0.3700 &  245     & 6.12      &   0.012  & T  &5.0 &35   &7,8  & CSS   &$<$0.06 &       & Gu\\ 
J0141+1353  & 3C49   & G & 0.621  &  134.0   & 6.71      &   0.007  & T  &2.31&21   &9,10 & CSS   &   1.17 &$-$185 & V \\ 
J0224+2750  & 3C67   & G & 0.310  &  25.8    & 11.3      &   0.012  & T  &2.56&0.5  &10,11& CSS   &$<$1.45 &       & V \\
J0251+4315  & S4     & Q & 1.311  &  449     & 0.11      &   0.008  & T  &1.38&13   &12   & CSS   &$<$2.61 &       & Gu\\
J0410+7656  & 4C76.03& G & 0.599  &  337     & 0.93      &   0.018  & T  &1.35&12   &10,13& CSS   &   2.66 &   315 & V \\ 
J0431+2037  & OF247  & G & 0.219  &  32.8    & 1.02      &   0.096  & T  &4.69&4.5  &13,14& GPS   &   3.66 &   318 & V \\ 
J0503+0203  & OG003  & Q & 0.585  &  196     & 0.07      &   0.028  & T  &1.39&24   &15   & GPS   &   6.69 &   43  & C\\
J0521+1638  & 3C138  & Q & 0.759  &  78.6    & 4.98      &   0.132  & T  &1.70&0.0  &9,10 & CSS   &$<$0.48 &       & V \\
            &        &   &        &          &           &          &    &    &     &16   &       &        &       &   \\ 
J0741+3112  & OI363  & Q & 0.635  &  238     & 0.04      &   0.279  & T  &1.26&2.9  &15,17& GPS   &$<$0.99 &       & V \\
J0954+7435  &        & G & 0.695  &  462     & 0.15      &$<$0.002  & D  &    &     &18,19& GPS   &$<$2.51 &       & V \\
J1035+5628  & OL553  & G & 0.459  &  72.0    & 0.02      &$<$0.002  & D  &    &     &20   & GPS   &$<$1.38 &       & V \\
J1120+1420  & 4C14.41& G & 0.362  &  37.7    & 0.42      &   0.017  & T  &1.28&15   &21   & GPS   &$<$0.61 &       & V \\
J1124+1919  & 3C258  & G & 0.165  &  3.0     & 0.28      &$<$0.012  & D  &    &     &11,22& CSS   &   5.05 &   163 & Gu\\ 
J1206+6413  & 3C268.3& G & 0.371  &  55.0    & 6.56      &   0.001  & T  &2.96&16   &23,10& CSS   &   2.07 &   258 & V \\
J1252+5634  & 3C277.1& Q & 0.321  &  26.8    & 7.74      &   0.068  & T  &2.53&0.0  &23,10& CSS   &$<$0.72 &       & V \\
J1308$-$0950& OP-010 & G & 0.464  &  166     & 3.47      &$<$0.010  & D  &    &     &1    & CSS   &$<$1.35 &       & V \\
J1313+5458  &        & Q & 0.613  &  72.0    & 0.25      &$<$0.004  & D  &    &     &18   & GPS   &$<$1.80 &       & V \\
J1326+3154  & 4C32.44& G & 0.370  &  88.7    & 0.29      &$<$0.005  & D  &    &     &24,25& GPS   &   0.75 &$-$471 & V \\
J1443+7707  & 3C303.1& G & 0.267  &  10.1    & 7.32      &$<$0.001  & D  &    &     &23   & CSS   &$<$1.50 &       & V \\
J1521+0430  & 4C04.51& Q & 1.296  &  1208    & 1.15      &$<$0.002  & D  &    &     &27   & GPS   &$<$0.64 &       & Gu\\
J1816+3457  &        & G & 0.245  &  5.60    & 0.15      &$<$0.004  & D  &    &     &19   & CSS   &   5.31 &$-$184 & P1\\
J1819$-$6345& MRC    & G & 0.0645 &  4.2     & 0.41      &   0.131  & T  &2.10&21   &1,28 & CSS   &   21.2 &$-$160 & M \\ 
J1821+3942  & 4C39.56& G & 0.798  &  319     & 3.88      &   0.098  & T  &2.75&5.5  &13,14& CSS   &   1.75 &$-$806 & V \\
J1944+5448  & OV573  & G & 0.263  &  17.3    & 0.16      &   0.033  & T  &1.81&5.0  &14,29& GPS   &   5.21 &$-$1420& V \\
J1945+7055  &        & G & 0.101  &  1.54    & 0.06      &   0.050  & T  &1.04&0.0  &30   & GPS   &   35.5 &$-$172&  P2\\
J2137$-$2042& OX-258 & G & 0.635  &  223     & 1.37      &$<$0.004  & D  &    &     &1,31 & CSS   &$<$1.18 &      &  V \\
J2325+4346  & OZ438  & G & 0.145  &  4.70    & 4.00      &$<$0.010  & D  &    &     &11   & CSS   &$<$0.84 &      &  Gu\\
J2344+8226  &        & Q & 0.735  &  292     & 1.94      &   0.134  & T  &1.39&11   &13,14& GPS   &$<$0.75 &      &  V \\
J2355+4950  & OZ488  & G & 0.238  &  22.3    & 0.26      &   0.011  & T  &1.03&7.0  &32,33& GPS   &   3.01 &$-$12 &  V \\
\hline
\end{tabular}
\end{center}
\begin{flushleft}
The columns are as follows: Cols. 1 and 2: source name and an 
alternative name; cols. 3 and 4: optical identification and redshift; col. 5: radio luminosity at 5 GHz; 
col. 6: the largest linear size (LLS) defined to be the separation between the outermost peaks of radio emission; 
col. 7: the fraction of emission from the core, f$_c$, at an emitted frequency of 8 GHz; 
col. 8: structural 
classification where D denotes a double source while T implies a triple source which has a detected core; 
col. 9: the separation ratio, R$_\theta$, defined to be the ratio of 
the separation of the further component from the core to 
the nearer one; col. 10: the misalignment angle, $\Delta$, defined to be the supplement of the angle formed at the
core by the outer hotspots; col. 11: references for the radio structure; col. 12: radio identification 
of the sources as CSS and GPS objects; col. 13: H{\sc i} column density or a 3$\sigma$ upper limit to it estimated 
assuming a spin temperature of 100 K and full coverage of the background source by the absorber; 
col. 14: the shift of the primary H{\sc i} component relative to the systemic velocity as measured from the optical 
emission lines, with a negative sign indicating a blue-shift, and col. 15: references for the H{\sc i} observations.

References for structural information:  1: Tzioumis et al. (2002); 2: Pearson \& Readhead (1988); 
3: Owsianik et al. (1998); 4: Zensus et al. (2002); 5: Baum et al. (1990); 6: Giovannini et al. (2001);
7: Feng et al. (2005); 8: Gupta et al. (2005); 9: Fanti et al. (1989); 10: Saikia et al. (1995); 
11: Sanghera et al. (1995); 12: Fey \& Charlot (2000); 13: Dallacasa et al. (1995); 14: Saikia et al. (2001); 
15: Stanghellini et al. (2001); 16: Akujor et al. (1991); 17: Saikia et al. (1998); 18: Taylor et al. (1994); 
19: Peck \& Taylor (2000); 20: Fomalont et al. (2000); 21: Bondi et al. (1998); 22: Strom et al. (1990); 
23: L\"udke et al. (1998); 24: Mutel et al. (1981); 25: Fey et al. (1996); 26: Perlman et al. (1996); 
27: Xiang et al. (2002); 28: Ojha et al. (2004); 29: Xu et al. (1995); 30: Taylor \& Vermeulen (1997); 
31: Fomalont et al. (2003); 32: Polatidis et al. (1995); 33: Taylor et al. (2000). \\ 
References for H{\sc i} observations:  V: Vermeulen et al. (2003); C: Carilli et al. (1998); Gu: Gupta et al. (2006); 
P1: Peck et al. (2000); M: Morganti et al. (2001); P2: Peck et al. (1999). \\
\end{flushleft}
\label{sample}
\end{table*}
%
%
\section{Introduction}
An understanding of the distribution and kinematics of the different components of the circumnuclear gas 
in an active galactic nucleus (AGN) is important both for studying the anisotropy in the radiation 
field and thereby testing the unification scheme (cf. Barthel 1989; Urry \& Padovani 1995), and for understanding 
the fuelling of the radio activity. At radio wavelengths, the ionized component of this gas may 
be probed via radio polarization measurements of compact components in sources which reside within the
dense interstellar environments of the parent galaxies. Such sources are the radio cores, and the compact 
steep-spectrum (CSS) and gigahertz peaked-spectrum (GPS) sources. The CSS sources are defined to be those with 
a projected linear size $\lapp$15 kpc (H$_o$=71 km s$^{-1}$ Mpc$^{-1}$, $\Omega_m$=0.27, $\Omega_\Lambda$=0.73, 
Spergel et al. 2003) and having a steep high-frequency radio spectrum 
($\alpha\gapp$0.5, where S($\nu)\propto\nu^{-\alpha}$). The GPS sources are more compact and have
sizes $\lapp$1 kpc (O'Dea 1998). 
An important way of probing the atomic gas on sub-galactic scales is by studying  21-cm H{\sc i} 
absorption towards the compact components of CSS and GPS objects, or the compact radio nuclei of the larger 
objects (e.g. van Gorkom et al. 1989; Conway \& Blanco 1995; Peck et al. 2000; Pihlstr\"om 2001; 
Vermeulen et al. 2003; Pihlstr\"om et al. 2003; Gupta et al. 2006).  

The H{\sc i} absorption lines associated with CSS and GPS objects exhibit a variety of line profiles, 
suggesting significant, sometimes complex, gas motions, while the  H{\sc i} column densities have been found to be 
anti-correlated with source sizes (Vermeulen et al.  2003; Pihlstr\"om et al. 2003; Gupta et al. 2006).  
Pihlstr\"om et al. have explored different distributions of the H{\sc i} gas density to explain this observed 
anti-correlation, and find that a disk as well as a 
spherical distribution are consistent with it.  
To study some of these aspects and to test for consistency with the unification scheme, we 
examine the dependence of H{\sc i} column density on relative prominence of the 
core, f$_c$, which is being used as a statistical measure of the orientation of the source axis to 
the line of sight (Orr \& Browne 1982; Kapahi \& Saikia 1982). 
In this paper, we have investigated this using a sample of 32 CSS and GPS sources, and have also examined
the dependence of H{\sc i} column density and 
velocity shift of the primary absorption component relative to the systemic velocity, on source symmetry parameters. 
\begin{figure}
\centerline{\vbox{
\psfig{figure=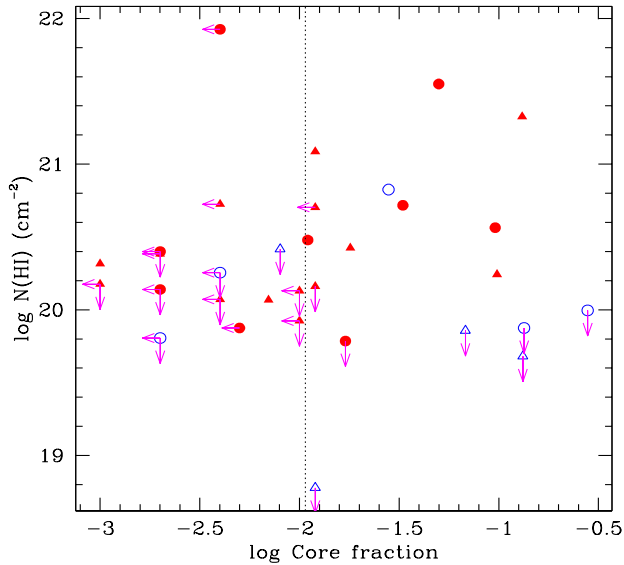,height=8.0cm,width=8.5cm,angle=0}
}}
\caption[]{Neutral hydrogen column density as a function of core fraction.
The solid symbols (in red) correspond to galaxies while open symbols (in blue) denote quasars.   
Circles denote GPS objects while triangles denote CSS sources. Upper limits are marked by arrows.  
The vertical dashed line marks the median value of f$_c$ for galaxies. 
} 
\label{nhivsfc}
\end{figure}

\section{The sample of sources}
Our sample consists of all CSS and GPS sources identified with radio galaxies and quasars for which we could 
compile information on the H{\sc i} column density and core flux density or an upper limit to it either from 
the literature or our observations. CSS and GPS sources such as J1347+1217 (4C+12.50) and J1407+2827 
(Mrk668, OQ208), which have prominent H{\sc i} absorption lines and well-defined radio structures 
have not been included because they are classifed as Seyferts in the literature (e.g. Hurt et al.
1999; Gallimore et al. 1999). The radio source J1415+1320 (OQ122) which has an H{\sc i} absorption
line (Carilli et al. 1992) but has been identified as a BL Lac object possibly residing in a spiral host galaxy 
(Wurtz, Stocke \& Yee 1996) has also not been considered. For this study, a core is defined to be a compact 
feature between the two outer lobes of emission. For those without a detected core, the 
upper limits to the core flux density are 3 times the rms noise in the image, and have been estimated only for 
those sources whose largest angular size is at least 5 times the resolution element.  
The fraction of emission from the core, f$_c$, or an upper limit to it at an emitted frequency of 8 GHz 
has been estimated for each source. To estimate this fraction, we have used a spectral index of 
0 and 1 for the core and extended emission respectively, unless information was available in the literature. 

The sample consisting of 32 objects of which 14 are GPS and 18 are CSS objects is listed in Table~\ref{sample}.
23 of these are associated with galaxies and 9 with quasars.  The projected linear size of the 
sample ranges from 0.02 to 11.3 kpc with a median value of 0.68 kpc, while the 5-GHz luminosity ranges 
from 1.3$\times$10$^{25}$ to 1.2$\times$10$^{28}$ W\,Hz$^{-1}$ with a median value of 7.6$\times$10$^{26}$ W\,Hz$^{-1}$. 
The values of f$_c$ range from $<$0.001 to 0.279 with a median value of 0.011, while those of N(H{\sc i}) range 
from $<$0.06$\times$10$^{20}$ to 84.2$\times$10$^{20}$ cm$^{-2}$ with a median value which is less than 
1.8$\times$10$^{20}$
cm$^{-2}$.  In this paper, we will refer to sources with core prominence lesser and greater than the median value as 
having lower and intermediate values. The sources with the highest values of f$_c$ are not represented in our sample 
since they would not be classified as CSS and GPS objects.   

%
%
\section{Dependence of H{\sc i} column density on core prominence}
To examine the dependence of H{\sc i} column density on core prominence and also check for consistency with 
the unification scheme for radio galaxies and quasars, we plot the H{\sc i} column density against the core 
fraction (Fig.~\ref{nhivsfc}).  In our sample, there is only one H{\sc i} detection for the 9 quasars and 
15 detections for the 23 galaxies, consistent with orientation effects playing a significant role in the 
detection of H{\sc i} absorption. 
The quasars have a higher median value of f$_c$ of $\sim$0.028 compared with 0.011 for the galaxies.  
These trends are broadly consistent with the expectations for the unification scheme.

One of the striking features seen in Fig.~\ref{nhivsfc} is that 10 of the 15 galaxies that have a detected H{\sc i} 
absorption feature, also have a detected radio core, compared with only 2 out 0f 8 sources without an H{\sc i}
detection. This suggests that detection of a radio core in galaxies enhances 
the probability of detecting a 21-cm absorption feature. Dividing the galaxies into two roughly equal groups at the 
median value of f$_c$$\sim$0.011, 9 of the 11 with f$_c$$\geq$0.011 (i.e. intermediate values of core fraction) have a 
detected absorption feature compared with 6 out of 12 detections in the other group with lower values of f$_c$. 
The median value of N(H{\sc i}), using upper limits as their values, for the intermediate f$_c$ group is 
$\sim$3.7$\times$10$^{20}$ cm$^{-2}$ compared with 1.5$\times$10$^{20}$ cm$^{-2}$ for the lower f$_c$ group. 
A Kolmogorov-Smirnov test shows the distributions to be different at a significance level of $\sim$95 per cent.
Thus for galaxies, there is a tendency for the detection rate as well as the column density to increase with the 
degree of core prominence.  For the 9 quasars in our sample, there is only one detection and hence it is not meaningful 
to do a similar exercise.  But clearly, sources with the lowest values of f$_c$ do not tend to have the higher 
values of N(H{\sc i}).

It is important to enquire whether such a trend is consistent with the unification scheme for radio galaxies 
and quasars.  The deficit of high column density values for sources with low values of f$_c$ can be
understood in a scenario where the radio source sizes are larger than the scale of the circumnuclear disk 
which contributes most of the absorbing gas, so that the line of sight to the lobes at very large inclinations 
(corresponding to small values of f$_c$) does not intersect the H{\sc i} disk.  For intermediate values of f$_c$, 
one would then expect higher detection rate as well as higher values of N(H{\sc i}) as seen in Fig.~\ref{nhivsfc}. 
For angles close to the line of sight, which would have the highest values of f$_c$, one would expect small 
values of N(H{\sc i}) and much lower detection rates due to the line of sight passing largely through the 
ionization cone and also through regions of the disk with a high spin temperature due to the presence of 
an AGN (Bahcall \& Ekers 1969).  Such sources will usually get classified as compact flat spectrum (CFS) sources 
and not many of them with similar properties as the sources in our sample have been searched for 
21-cm absorption (cf. Gupta et al. 2006).

\subsection{Dependence on linear size}
\begin{figure}
\centerline{\vbox{
\psfig{figure=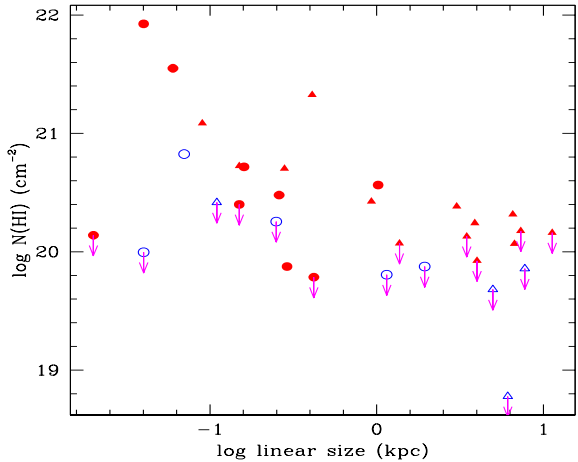,height=6.5cm,width=8.0cm,angle=0}
}}
\caption[]{Neutral hydrogen column density as a function of projected linear size of the radio sources.  
See caption of Fig.~\ref{nhivsfc} for the meaning of the various colours and symbols.
} 
\label{nhivsl}
\end{figure}
%
The neutral hydrogen column density as a function of linear size for our sample of 32 radio sources is 
shown in Fig.~\ref{nhivsl}. This shows the anti-correlation between the H{\sc i} column density and source 
size reported by Pihlstr\"om et al. (2003) and extended to a larger sample of 96 sources by Gupta et al. (2006). 
Thus detection probability of 21-cm absorption feature is higher for both small sources and those with intermediate 
values of core prominence.  It may be noted that projected linear size and f$_c$ for this sample show no significant relation.

\begin{figure}
\centerline{\vbox{
\psfig{figure=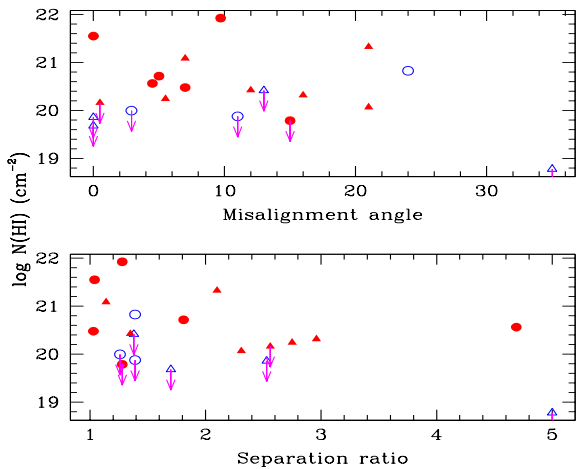,height=6.5cm,width=8.0cm,angle=0}
}}
\caption[]{Neutral hydrogen column density as a function of radio source symmetry paramters i.e. 
the separation ratio and misalignment angle.
See caption of Fig.~\ref{nhivsfc} for the meaning of the various colours and symbols.
} 
\label{nhivsasym}
\end{figure}
\section{H{\sc i} and symmetry parameters}
\begin{figure}
\centerline{\vbox{
\psfig{figure=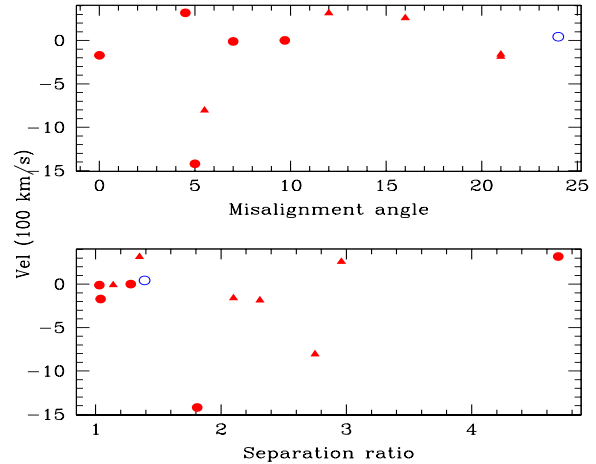,height=6.5cm,width=8.0cm,angle=0}
}}
\caption[]{
The plot of velocities for the principal H{\sc i} absorption components with respect to the
systemic velocity estimated from optical lines as a function of symmetry parameters.
See caption of Fig.~\ref{nhivsfc} for the meaning of the various colours and symbols.
} 
\label{vshvsasym}
\end{figure}

In addition to the circumnuclear disk, absorption could also occur in neutral clouds that are part of the gas 
that has been accelerated by interaction with a radio jet.  Gas outflows as fast and broad as 
$\sim$1000 km s$^{-1}$ have been detected in a number of radio sources in neutral as well as ionised gas 
(see Morganti, Tadhunter \& Oosterloo 2005).  The kinematical and spatial properties of these suggest that 
they are being driven by the same mechanism i.e. interaction with the expanding radio jets.  
The dynamical interaction of the radio source with these clouds could distort the morphology of the radio source 
thereby affecting the symmetry parameters such as the separation ratio and the misalignment angle. 
The separation ratio, R$_\theta$, is defined to be the ratio of 
the separation of the further component from the core to 
the nearer one, while the misalignment angle, $\Delta$, is defined to be the supplement of the angle formed at the
core by the outer hotspots. 
Such jet-cloud interactions may also affect the velocity of the absorbing cloud relative to the systemic 
velocity of the galaxy. However, we find no evidence of any significant dependence of the H{\sc i} column density 
as well as the relative velocity of the primary absorption component on either the separation ratio or the 
misalignment angle (Figs.~\ref{nhivsasym} and  \ref{vshvsasym}). 
But this does not rule out the importance of the role played by jet-cloud interactions in the observed structures of
these sources as 
there exists significant evidence for this in terms of structural, brightness and polarization 
symmetry parameters (Saikia et al. 1995; 2001; Saikia \& Gupta 2003).  

%
%
%
\section {Summary}
For a sample of CSS and GPS objects, we have examined the dependence of 
H{\sc i} column density on the degree of core prominence and projected linear size. 
 In our sample, 15 out of 23
galaxies exhibit 21-cm absorption as against 1 out of 9 quasars, which is broadly 
consistent with the unification scheme. 
The detection rate as well as the H{\sc i} column density appears to be larger for galaxies 
with more prominent cores, 
with the higher column densities occurring in intermediate values of f$_c$ and not in the lowest values of f$_c$.  
This broad trend can be understood in the unification scheme if the lines of sight towards the lobes of sources with 
the lowest values of f$_c$, which are inclined at the largest angles to the line of sight, do not pass through the 
circumnuclear H{\sc i} disk.  This could happen if the source sizes are larger than the scale of the disk. 
It is interesting to note that 10 of the 15 galaxies with a detected H{\sc i} absorption 
feature also has a detected radio core, compared with 2 out of 8 galaxies without any H{\sc i} detection. 
Our sample also shows an inverse relationship between the 
H{\sc i} column density and the projected linear size, suggesting that both small sizes and more prominent cores 
enhance the probability of detecting H{\sc i} absorption in CSS and GPS objects.

In addition to the disk, H{\sc i} absorption may also be due to clouds of cold gas, some of which may
have interacted with the radio jet. We, however, find that the H{\sc i} column density and the velocity shift of 
the primary absorption component relative to the systemic velocity show no dependence on the degree of misalignment 
and the separation ratio. 
 
%
%
\section*{Acknowledgments}
We thank an anonymous referee and Ian Browne for several useful comments and suggestions. We 
also thank the numerous contributers to the GNU/Linux group. This research has made use of 
the NASA/IPAC Extragalactic Database (NED) which is operated by the Jet Propulsion Laboratory, 
California Institute of Technology, under contract with the National Aeronautics and Space 
Administration. DJS thanks dear Rowena and Neeraj, Chiranjib and Ananda for either encouraging or 
helping to move along through work and writing. DJS also thanks Andrew Lyne, Director, Jodrell Bank 
Observatory, and Peter Thomasson for hospitality where part of the work was done.

%

\label{lastpage}

\end{document}